\documentclass[english,aps,prl,twocolumn,amsmath,amssymb,showpacs,superscriptaddress,notitlepage,longbibliography]{revtex4-2}
\usepackage{bm}
\usepackage[pdftex]{graphicx,hyperref}
\hypersetup{colorlinks = true, urlcolor = blue, linkcolor = blue, citecolor = blue}
\usepackage{mathtools}
\usepackage{ulem}
\usepackage{color}

\begin{document}

\title{Vortex States and Coherence Lengths in Flat-Band Superconductors}

\author{Chuang Li}
\affiliation{Center for Correlated Matter and School of Physics, Zhejiang University, Hangzhou 310058, China}
\affiliation{Hefei National Laboratory, Hefei 230088, China}
\author{Fu-Chun Zhang}
\email{fuchun@ucas.ac.cn}
\affiliation{Kavli Institute for Theoretical Sciences, University of Chinese Academy of Sciences, Beijing 100190, China}
\affiliation{Collaborative Innovation Center of Advanced Microstructures, Nanjing University, Nanjing 210093, China}
\author{Lun-Hui Hu}
\email{lunhui@zju.edu.cn}
\affiliation{Center for Correlated Matter and School of Physics, Zhejiang University, Hangzhou 310058, China}

\begin{abstract}
Superconductivity in flat-band systems, governed by quantum metric of Bloch states rather than the BCS framework, exhibits unique phenomena due to the vanishing electron group velocity. 
%Here, we present a comprehensive study of vortex states and coherence lengths in flat-band superconductors, and reveal how quantum geometry determines their fundamental properties. 
Here, we propose the vortex states and vortex size as direct probes to explore the quantum geometry effects in flat-band superconductors. 
We show that flat-band vortex bound states are sharply localized near the vortex core, and the energy gap between the lowest two bound states is on the order of the bulk superconducting gap. Both the spatial spread and energy scales of bound states are controlled by the flat-band's quantum metric length. Moreover, the vortex size at zero temperature, set by the quantum metric length, is atomic in scale and independent of interaction strength. Near $T_c$, the vortex size corresponds to the Ginzburg-Landau coherence length, diverges as $\xi\sim \sqrt{T_c/(T_c-T)}\xi_0$, where $\xi_0$ depends linearly on the quantum metric length. Thus, the quantum metric serves as the lower bound for vortex state spread and vortex size. We also introduce perturbations to make the flat band dispersive, and distinguish flat-band vortices from BCS-like vortices. Our results establish vortices as universal probes of quantum geometry in flat-band superconductors.
\end{abstract}

\maketitle

\textit{\color{blue}Introduction.--}
In solid-state physics, quantum geometry, encoded in the quantum geometric tensor of Bloch states, governs fundamental electronic and topological properties~\cite{Resta2011EPJB,Paivi2023PRL,XXinChen2024NSR,YJiaBin2025arXiv}. 
The quantum geometric tensor, whose imaginary part gives the Berry curvature~\cite{Berry1984RSPA,Simon1983PRL,Aharonov1987PRL}, plays a fundamental role in determining band topology for free electrons in quantum materials~\cite{Thouless1982PRL,Kane2010RMP,QXiaoLiang2011RMP,CChingKai2016RMP,Bernevig2017Nat,Vishwanath2017NatCom,Kruthoff2017PRX}. 
Recent advances have revealed that its real part, the quantum metric~\cite{Provost1980CMP}, has become equally crucial for understanding both non-interacting and correlated systems~\cite{David2012RMP,Rossi2021COSSMS,Paivi2022NRP}. 
In free-electron quantum materials, it can determine transport phenomena such as optical responses~\cite{Vishwanath2022NatPhys} and Landau level quantization~\cite{Hwang2021NatCom}, while in interacting systems, it underpins emergent states like flat-band superconductivity~\cite{Kopnin2011PRB,Paivi2015NatCom,Paivi2016PRL,Paivi2017PRB,LBiao2019PRL,Balents2020NatPhys,Paivi2022PRB,JGuoDong2025PNAS}, 
exciton condensates~\cite{HXiang2022PRB,Queiroz2024PRL}, unconventional superconductivity~\cite{Kitamura2024PRL,HWen2024PRL}, and electron-phonon coupling strength~\cite{Bernevig2024NatPhys,ZhPengHao2024PRB}. Moreover, the quantum metric’s contribution to key properties like the superfluid stiffness in flat-band superconductors has been experimentally demonstrated, including nonlinear transport measurements~\cite{Bockrath2023Nat} and microwave circuit quantum electrodynamics~\cite{William2025Nat} in twisted bilayer graphene, and radio-frequency reflectometry techniques in trilayer graphene~\cite{banerjee2025superfluid}, distinguishing it from BCS superconductors.

Correlated states are determined by both electronic dispersion and quantum geometry of Bloch bands, with the latter playing the dominant role in flat-band systems. For instance, the flat-band superconductors are well defined when the bandwidth at the Fermi level is much smaller than the superconducting gap $\Delta_0$~\cite{Kivelson2024PRX}. 
If the flat band is energetically isolated from other bands by a gap far exceeding $\Delta_0$, the system is in the isolated flat-band limit. In this scenario, based on the BCS framework, such extreme band flatness implies a vanishing superconducting coherence length $\xi_{\text{BCS}} \approx v_F/\Delta_0$, as the electronic group velocity $v_F$ approaches zero. However, recent studies reveal that both zero-temperature pair correlations and Ginzburg-Landau (GL) theory near $T_c$ predict a finite coherence length in the isolated flat-band superconductor~\cite{Law2024PRL,Iskin2023PRB,Iskin2024PRB,Law2025ComPhys,Maxime2025SciPost}.
The GL coherence length is experimentally measurable, and is predicted as $\xi_{\text{GL}}(T)=\sqrt{T_{\text{MF}}/(T_{\text{MF}}-T)}[\det(\bar{g}_{ab})]^{1/4}$, where $\bar{g}_{ab}$ denotes the quantum metric length that is a spatial scale derived from the Brillouin-zone-averaged quantum metric~\cite{David1997PRB}. %Consequently, this provides a novel experimental way to directly probe the quantum metric as the fundamental driver of flat-band superconductors.

% Emphasize measurement by STM rather than Hc2
Despite progress in understanding superfluid stiffness and coherence length in flat-band superconductors, the physics of Abrikosov vortices~\cite{Caroli1964PL,Gygi1991PRB,Hayashi1998PRL}, a hallmark of type-II superconductivity~\cite{Abrikosov2004RMP,Tinkham2004book}, remains largely unexplored. Crucially, scanning tunneling microscopy can directly resolve the spatial profile of $\Delta_0$ with atomic-scale precision~\cite{KLingYuan2025Nat}, enabling vortices to sever as direct probes of quantum geometric effects through measurements of: (i) vortex bound-state energy spectra, (ii) localized wavefunctions, and (iii) vortex core sizes or $\xi_{\text{GL}}$. Furthermore, the proposed connection between the GL coherence length and the quantum metric length can be rigorously tested through tight-binding model calculations at both zero temperature and near $T_c$, circumventing the limitations of the GL framework.

\begin{figure*}[!htbp]
\centering
\includegraphics[width=0.75\linewidth]{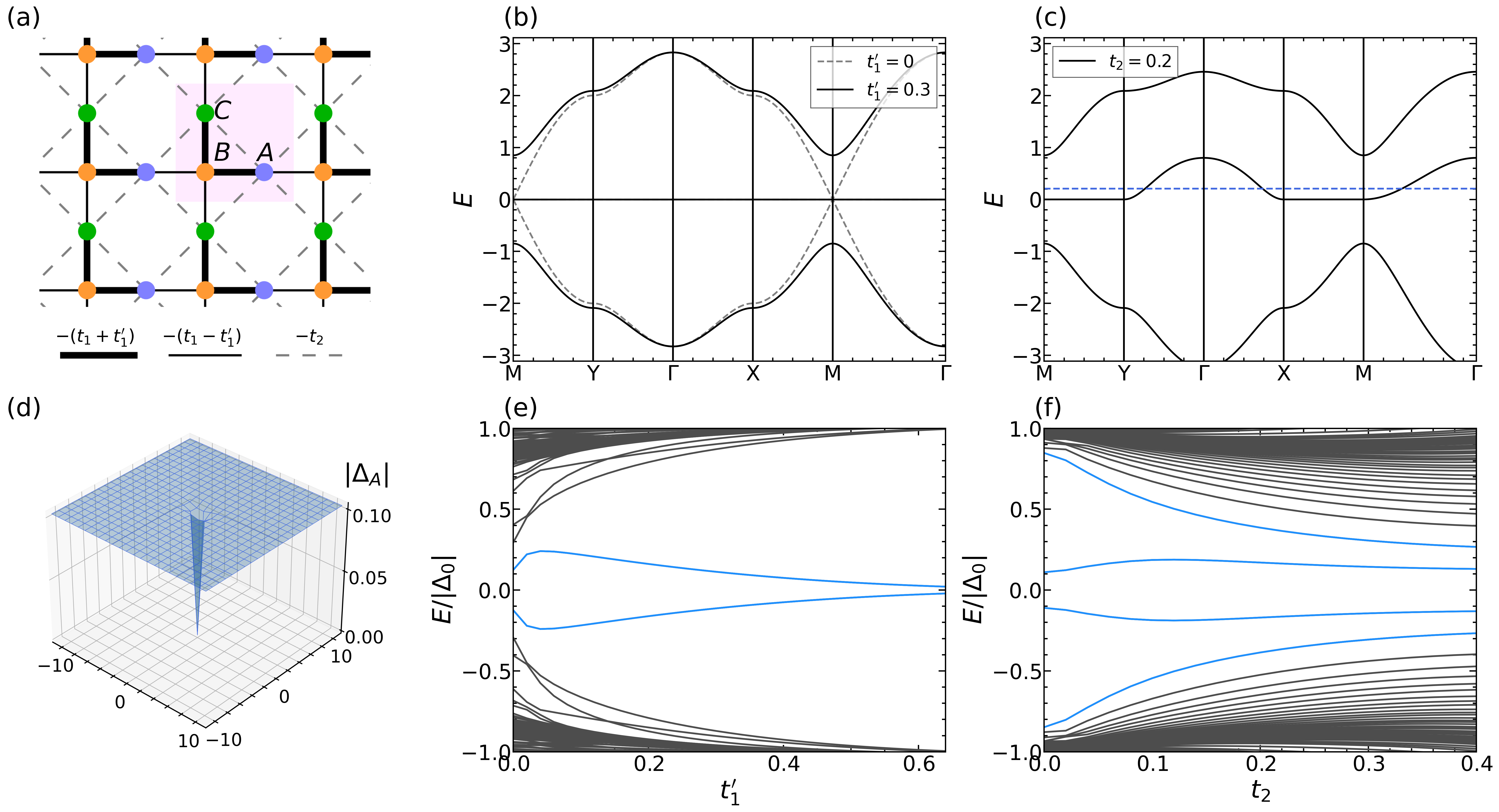}
\caption{(a) Schematic of the Lieb lattice structure, with the unit cell outlined by the light magenta rectangle. 
(b) Band structures with strictly flat band in the Lieb lattice model without band gap ($t_1'=0$, dashed gray line) and with band gap ($t_1'=0.3$, black solid line).
(c) Band structures for $t_1'=0.3$ and $t_2=0.2$. The middle band becomes dispersive. The dashed blue line indicates the Fermi energy at half-filling.
(d) Spatial profile of $\Delta_A(\mathbf{r})$ near the vortex core in the flat-band vortex limit at zero temperature. 
(e) Evolution of vortex-bound-state spectrum with $t_1'$ at fixed $t_2=0$. $t_1' \gg 0.04$ reaches the isolated flat-band vortex limit.
(f) Evolution of vortex-bound-state spectrum with $t_2$ at fixed $t_1'=0.3$. $t_2 \ll 0.025$ reaches the isolated flat-band vortex limit.}
\label{fig1}
\end{figure*}

In this work, we investigate superconducting vortices in flat-band systems and reveal how quantum geometry governs their fundamental properties. We find that the vortex bound states are sharply localized around the vortex core, with their spatial spread and the corresponding energy controlled by the flat-band quantum metric. Notably, the energy gap between the two lowest in-gap bound states is on the order of $\Delta_0$, serving as a smoking gun for experimental detection of flat-band vortices. At zero temperature, the vortex size remains atomic-scale (of order a lattice constant) and independent of both $\Delta_0$ and interaction strength, which can be distinguished from BCS vortices. Near $T_c$, the vortex size corresponds to the $\xi_{\text{GL}}$ and follows Ginzburg-Landau scaling $\xi\sim \sqrt{T_c/(T_c-T)}\xi_0$, where $\xi_0$ depends linearly on the quantum metric length $\bar{g}_{ab}$ near $T_c$. Therefore, our numerical results indicate that the quantum metric serves as the universal lower bound for characteristic scales in flat-band vortices.

\textit{\color{blue}Flat-band quantum metric.--}
We study a superconducting system where the chemical potential lies exactly within a flat band with quantum metric. The general mean-field Hamiltonian is give by
\begin{align} \label{eq:hmtn-tot}
H_{sc} = H_{fb} -\mu \sum_\alpha c_\alpha^\dagger c_\alpha +H_{pair},
\end{align}
where $H_{fb}$ describes a flat-band system with quantum metric, $\mu$ denotes the chemical potential, $c_\alpha$ is the electron annihilation operator acting on the $\alpha$ orbital, and $H_{pair}$ is the pairing potential. As a concrete example, we focus on the Lieb lattice model, whose normal-state Hamiltonian in momentum space is
\begin{align} \label{eq:hmtn-Lieb}
H_{fb}(\mathbf{k}) = C^\dagger \begin{pmatrix}
    0 & h_{AB} & h_{AC} \\
    h_{AB}^* & 0 & h_{BC} \\
    h_{AC}^* & h_{BC}^* & 0
\end{pmatrix} C \ ,
\end{align}
where the basis is defined as $C=(c_A, c_B, c_C)^T$. The unit cell comprises three sublattices, labeled by A, B and C as shown in Fig.~\ref{fig1}(a). The nearest-neighbor hopping terms in the unit-cell gauge are given by: $h_{AB} = -(t_1+t_1')-(t_1-t_1')e^{ik_x}$ and $h_{BC} = -(t_1+t_1')-(t_1-t_1')e^{-ik_y}$, where $-(t_1+t_1')$ [thick solid lines in Fig.~\ref{fig1}(a)] and $-(t_1-t_1')$ [thin solid lines in Fig.~\ref{fig1}(a)] represent the intra-cell and inter-cell hoppings respectively. We set $t_1=1$ as energy unit. The parameter $t_1'$ determines the band gap $\Delta E_{M}=2\sqrt{2} t_1'$ between the flat band and the other two dispersive bands at the $M$ point. When $t_1'=0$, the dispersive bands touch the flat band at the $M$ point, while $t_1'\neq0$ opens a gap and isolates the flat band [Fig.~\ref{fig1}(b)]. Thus, the flat band becomes isolated simply by increasing $|t_1'|$. The quantum metric of the flat band is given by 
\begin{align}
g_{ab}(\mathbf{k})=\text{Re}[\langle \partial_a \psi_0 | \partial_b \psi_0 \rangle - \langle \partial_a \psi_0 |\psi_0 \rangle \langle \psi_0 | \partial_b \psi_0 \rangle] ,
\end{align}
where $\psi_0(\mathbf{k})$ the flat band eigenstate of $H_{fb}(\mathbf{k})$, $\partial_a \equiv \partial /\partial k_a$ and $a,b\in\{x,y\}$. $g_{ab}(\mathbf{k})$ of the isolated flat band exhibits $t_1'$-dependent behavior, with its dominant contributions localized near the $M$ point [see Supplementary material (SM)~\cite{SP} for detailed analysis].
In particular, $g_{xx}(\pi,\pi)=-g_{xy}(\pi,\pi)=[(t_1-t_1')/4t_1']^2$ demonstrating an inverse quadratic scaling with $t_1'$. In addition, the next-nearest-neighbor hopping term is $h_{AC} = -t_2 (1+e^{ik_x}) (1+e^{-ik_y})$, represented by gray dashed lines in Fig.~\ref{fig1}(a). It introduces dispersion to the flat band when $k_x\neq \pm\pi$ and $k_y \neq \pm\pi$, as shown in Fig.~\ref{fig1}(e) for $t_2 = 0.2$, where the orange dashed line marks the system's half-filling condition. The flat band develops a finite bandwidth $W_{fb}=4|t_2|$.

\textit{\color{blue}Flat-band vortex states.--}
To induce superconductivity in the flat band, we construct the pairing Hamiltonian $H_{pair}$ through mean-field decoupling of the attractive Hubbard interaction $H_\text{int} = -U \sum_{\mathbf{r},\alpha} \hat{n}_{\mathbf{r}\alpha\uparrow} \hat{n}_{\mathbf{r}\alpha\downarrow}$ with $U>0$, where $\hat{n}_{\mathbf{r}\alpha\uparrow}$ and $\hat{n}_{\mathbf{r}\alpha\downarrow}$ are electron density operators on sublattice $\alpha$ in unit cell $\mathbf{r}$.
%This mean-field framework holds for moderate $t_1'$ but breaks down for large $t_1'$~\cite{Paivi2016PRL}. 
In this work, we investigate vortex states in flat-band superconductors. The vortex configuration is encoded in the pairing term
\begin{align}
H_{pair} = \sum_{\mathbf{r},\alpha} \left(\Delta_{\alpha}(\mathbf{r}) e^{i\varphi(\mathbf{r})} c_{\mathbf{r}\alpha\uparrow}^\dagger c_{\mathbf{r}\alpha\downarrow}^\dagger +\text{H.c.} \right),
\end{align}
where $e^{i\varphi(\mathbf{r})}$ represents the vortex phase winding for on-site $s$-wave Cooper pairs, with $\varphi(\mathbf{r})$ denoting the azimuthal angle of unit-cell position $\mathbf{r}$ relative to the vortex core~\footnote{For the unit cell with $r=0$, we set $\Delta_A=\Delta_B=\Delta_C=0$}. The spatial profile of the gap function $\Delta_\alpha(\mathbf{r})$ on sublattice $\alpha$ is determined self-consistently through the gap equation: $\Delta_{\alpha}(\mathbf{r}) e^{i\varphi(\mathbf{r})} = -U \langle c_{\alpha\downarrow}(\mathbf{r}) c_{\alpha\uparrow}(\mathbf{r}) \rangle$.
%which accounts for the local pairing correlations. 
The vortex core is positioned at the central unit cell of the system. Our calculations reveal that the flat-band vortex states are sharply localized, rendering finite-size effects negligible for the $31\times31$ unit cell. Since the electron density of the flat band is concentrated on the A and C sublattices, the local pairing follows $\Delta_A=\Delta_C \gg \Delta_B$~\cite{Paivi2016PRL}. Figure~\ref{fig1}(d) displays $\Delta_A(\mathbf{r})$ near the vortex core, showing its suppression to zero at the core and recovery to the bulk gap $\Delta_{0,A}$ away from the core.

To study properties of flat-band vortices, we first identify the parameter region for an isolated flat-band superconductor that requires $ W_{fb} \ll \Delta_{0,A} \ll \Delta E_M $. This constrains the hopping parameters to $|t_1'| \gg \Delta_{0,A}/2\sqrt{2}$ and $|t_2| \ll \Delta_{0,A}/4$. Therefore, this parameter space enables systematic exploration of two distinct crossovers: (i) from isolated to non-isolated flat-band superconductor, and (ii) from flat-band to BCS-type superconductor. Substituting the self-consistently calculated $\Delta_{\alpha}(\mathbf{r})$ into Eq.~\eqref{eq:hmtn-tot}, we calculate the spectra of the vortex systems, and show the results of crossover (i) in Fig.~\ref{fig1}(e). In particular, at fixed $U=0.4$ and $t_2=0$, we obtain $\Delta_{0,A} \approx 0.1$ and $\Delta_{0,B} \approx 0.005$ which is almost irrelevant to changing $t_1'$. In Fig.~\ref{fig1}(e), $t_1' \approx 0.04$ reflects the turning point observed from the energy of the lowest vortex state. This marks the critical $t_1'$ where the vortex bound states that originate from the other dispersive bands merge into the bulk continuum, thereby achieving the isolated flat-band vortex after increasing $t_1'$. More details of the projected $B$-sublattice wavefunction that stems from the dispersive bands at this turning point is presented in SM~\cite{SP}.

To further elucidate distinctive features of flat-band vortices in comparison to BCS vortices, we introduce finite $W_{fb}$ to study the crossover behavior (ii). In general, BCS vortices typically host bound states numbering approximately $\mu/|\Delta_0|$~\cite{Caroli1964PL}, the $\mu\to0$ limit in flat-band systems suggests only order of one vortex state should persist. This picture is consistent with our numerical results, where we set $t_1'=0.3$ while increasing $t_2$ from $0$ to $0.4$. By adjusting $U$, we can maintain both the pairing gap and the half-filling condition (i.e.,~$\mu \approx t_2$) as $t_2$ changes~\footnote{The required interaction strength $U$ can be estimated by using the gap profile $\Delta_\alpha=\{\Delta_0,0,\Delta_0\}$, where we neglect the small $\Delta_{0,B}$.}. As shown in Fig.~\ref{fig1}(f), the number of vortex states increases significantly as $t_2 \gg 0.025$, and the energy splitting between the lowest two bound states exhibits linear $1/t_2$ dependence.
%\textcolor{red}{LH: one more sentence.}
%\textcolor{cyan}{Lc: In Fig.~\ref{fig1}(f), when $t_2 \gg 0.025$, the isolated flat-band vortex states evolve continuously as part of discrete CdGM states.}

\begin{figure}[t]
\centering
\includegraphics[width=\linewidth]{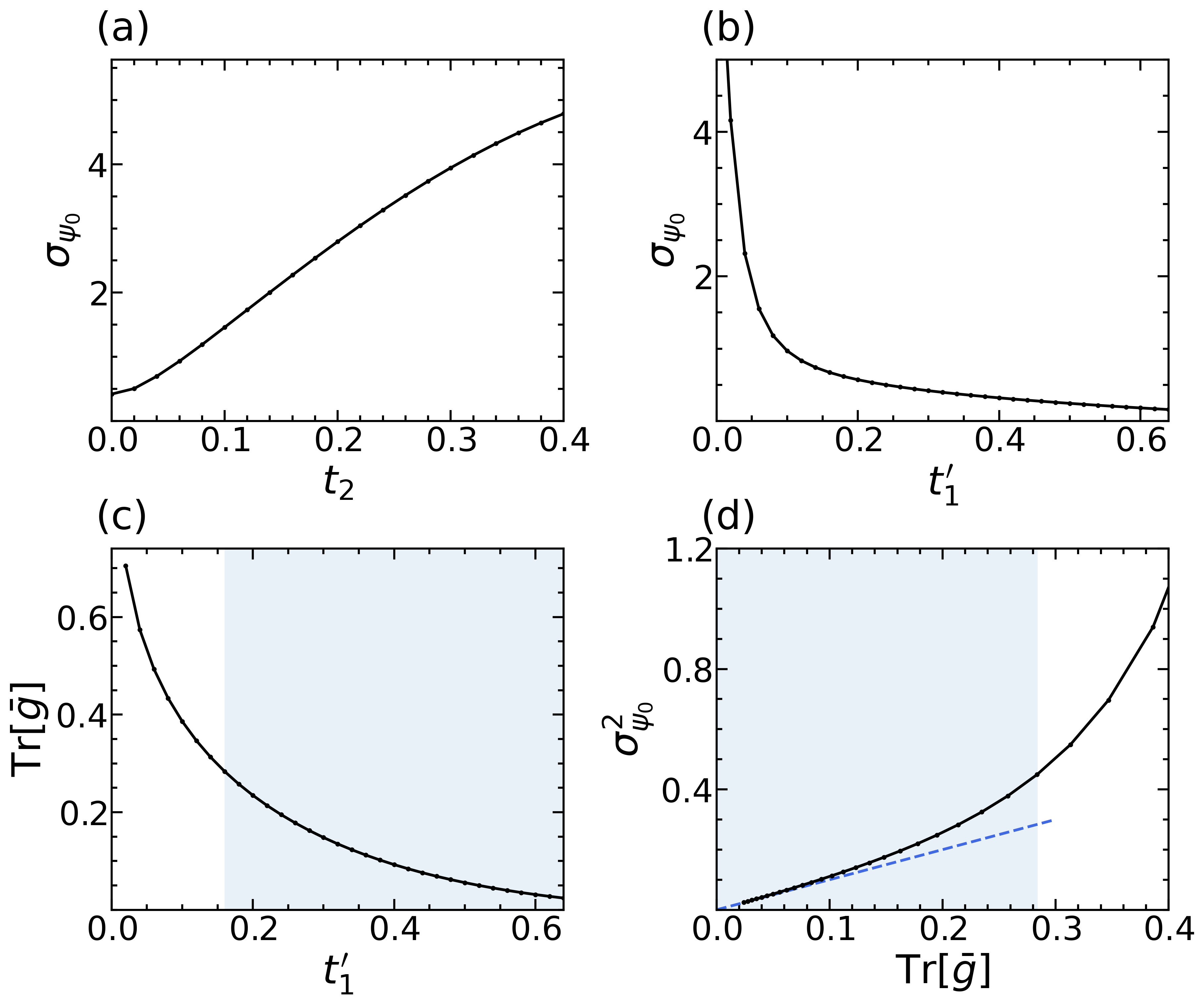}
\caption{(a) The wavefunction spread $\sigma_{\psi_0}$ for the lowest-energy state $\vert\psi_0\rangle$, depends on the band-flattening parameter $t_2$ (fixed $t_1'=0.3$). 
(b) Evolution of $\sigma_{\psi_0}$ with the band-gap parameter $t_1'$ (fixed $t_2=0$).
(c) Trace of the average quantum metric $\bar{g}$ as a function of $t_1'$. 
(d) Quadratic scaling relation between $\sigma_{\psi_0}^2$ and $\text{Tr}[\bar{g}]$. The dash line indicates $\sigma_{\psi_0}^2=\text{Tr}[\bar{g}]$.
Light blue region denotes the isolated flat-band vortex region.
}
\label{fig2}
\end{figure}

%The lowest energy state is a vortex bound state and decays rapidly without oscillation away from the vortex core.

Notably, Figs.~\ref{fig1}(e) and \ref{fig1}(f) both reveal that the lowest vortex bound state's energy decreases monotonically as the system approaches the isolated flat-band vortex limit (i.e.,~with increasing $t_1'$ or decreasing $t_2$). This energy scaling reflects the state's strongly localized character, evidenced by its rapidly decaying, oscillation-free wave function away from the vortex core. To quantify this behavior, we compute the standard deviation of the position operator, $\sigma_{\psi_0} = \sqrt{\langle r^2 \rangle -\langle \mathbf{r} \rangle^2}$, which measures the bound state's spread. Here, we focus on the lowest energy vortex state $\psi_0$, and provide the results for other higher-energy states in SM~\cite{SP}.
As shown in Fig.~\ref{fig2}(a), $\sigma_{\psi_0}$ increases during the crossover from the isolated flat-band vortex to the BCS vortex state as $t_2$ is enhanced. Similarly, in the isolated flat-band vortex region ($t_1' \gg 0.04$), the flat-band vortex state's energy shows a positive relationship with $\sigma_{\psi_0}$, as shown in Fig.~\ref{fig2}(b). These results indicate a direct connection between energy scaling and wave function spread for flat-band vortex: $E_{\psi_0} \propto \sigma_{\psi_0}$.

However, this relationship breaks down when $t_1' < 0.04$ due to hybridization between dispersive-band and flat-band vortex states [Fig.~\ref{fig1}(b)]. In this regime, the dispersive-band vortex states can become lower in energy than that of the flat-band vortex because of $\Delta_{0,B} \ll \Delta_{0,A/C}$. This hybridization accounts for the observed non-monotonic dependence of $E_{\psi_0}$ on $\sigma_{\psi_0}$, as seen by comparing Fig.~\ref{fig1}(e) and Fig.~\ref{fig2}(b). More details are provided in SM~\cite{SP}.
%\textcolor{cyan}{Lc: B-sublattice has a vortex with larger size and smaller $\Delta_{0,B}$, this may allow the eigenstates with large B component to have small energy.}

It is noteworthy that the spread of the lowest-energy isolated flat-band vortex state is governed by the parameter $t_1'$. This parameter \textit{solely} modulates the quantum metric length scale of flat-band electrons, as demonstrated by the monotonic decrease of $\text{Tr}[\bar{g}]$ with increasing $t_1'$ in Fig.~\ref{fig2}(c). Here, $\bar{g}_{ab}$ denotes the average of $g_{ab}(\mathbf{k})$ over the first Brillouin zone~\cite{David1997PRB}. This suggests a direct correspondence between $\sigma_{\psi_0}$ and $\bar{g}_{ab}$. Figure~\ref{fig2}(d) explicitly compares $\sigma_{\psi_0}^2$ with $\text{Tr}[\bar{g}]$, as both share the same dimension. We find that: (i) $\sigma_{\psi_0}^2$ scales linearly with $\text{Tr}[\bar{g}]$ ($\sigma_{\psi_0}^2 \propto \text{Tr}[\bar{g}]$)
%\textcolor{cyan}{(Lc: $\sigma_{\psi_0}^2$ is positively associated with $\text{Tr}[\bar{g}]$)} 
across the parameter space; and (ii) $\text{Tr}[\bar{g}]$  serves as an asymptotic lower bound for $\sigma_{\psi_0}^2$ as $\text{Tr}[\bar{g}] \to 0$.
%$\text{Tr}[\bar{g}]$ can provide a fundamental lower bound for $\sigma_{\psi_0}^2$. %\textcolor{cyan}{(Lc: $\text{Tr}[\bar{g}]$ becomes an asymptotic lower bound for $\sigma_{\psi_0}^2$ as $\bar{g} \to 0$)} 
Therefore, this inequality establishes the quantum metric length $\bar{g}_{ab}$ as determining the minimal spatial spread of flat-band vortex states.

\begin{figure}[t]
\centering
\includegraphics[width=\linewidth]{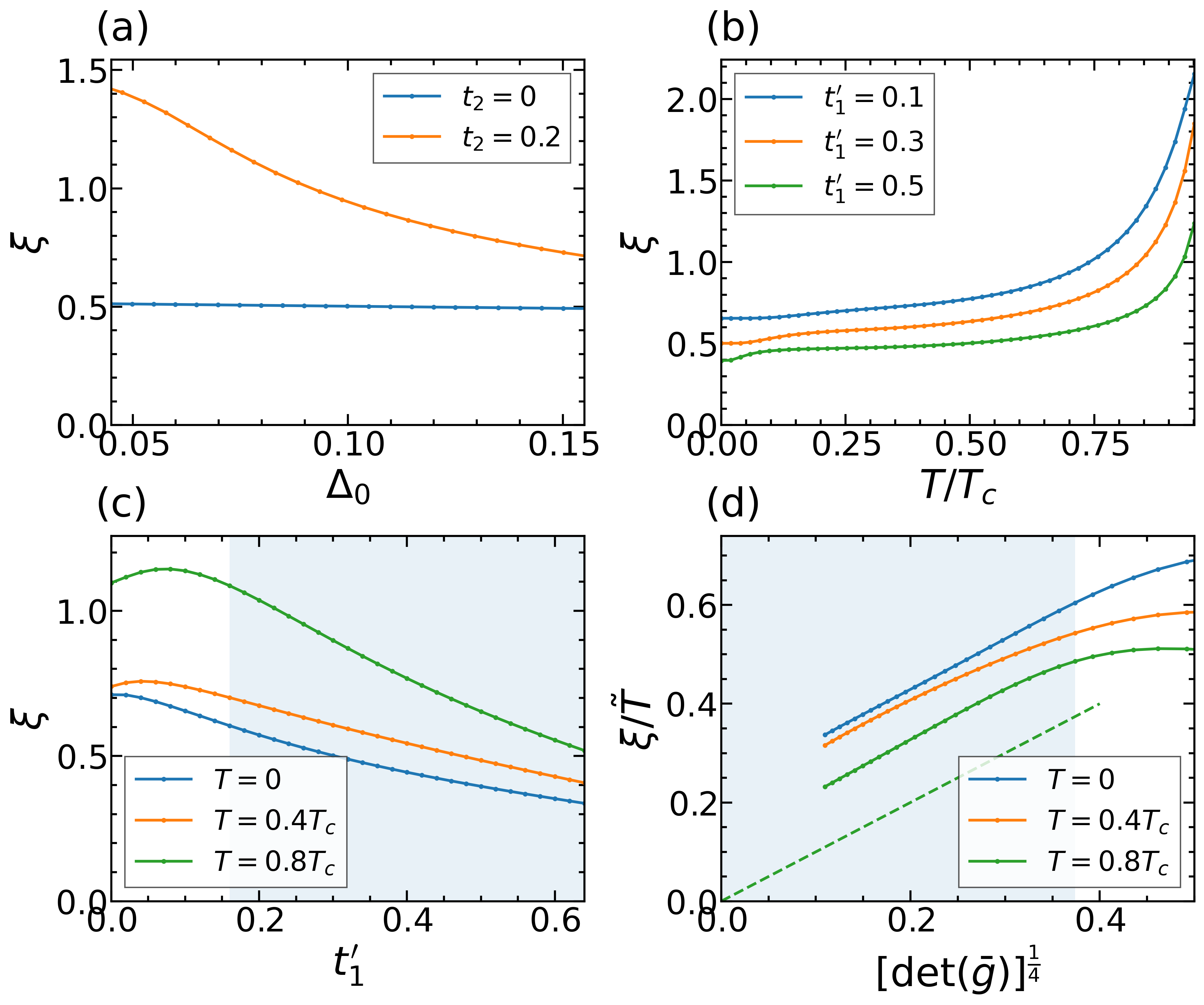}
\caption{(a) The vortex size $\xi(T=0)$ versus superconducting gap $|\Delta_0|$ for isolated flat-band vortices ($t_2=0$, red line) and BCS-like vortices ($t_2\approx \mu \gg |\Delta_0|$, blue line). 
(b) Temperature evolution of $\xi(T)$ for the flat-band case at different $t_1'$, showing universal critical divergence as $T\to T_c$.
(c) Dependence of $\xi$ on quantum metric parameter $t_1'$ across different temperatures.
(d) Universal scaling of $\xi/\tilde{T}$ with quantum metric length $[\det{\bar{g}}]^{1/4}$ at different temperatures. The dash green line corresponds to the GL coherence length $\xi_{\text{GL}}/\tilde{T}=[\det{\bar{g}}]^{1/4}$ from Eq.~\eqref{eq:xi-det-g}.
Light blue region denotes the isolated flat-band vortex region.
}
\label{fig3}
\end{figure}

\textit{\color{blue}Flat-band vortex size.--}
We next examine the spatial structure of $\Delta_\alpha$ in a flat-band vortex, establishing a experiment accessible method to quantify the quantum metric effect. The vortex radius is determined by extracting the magnitude of $\Delta_\alpha$ near the vortex core along a specific direction and fitting to a $\tanh$ function~\cite{Tinkham2004book},
\begin{align}
|\Delta_\alpha(\mathbf{r})|=|\Delta_{0,\alpha}| \tanh(r/\xi_{\alpha}) ,
\end{align}
where $|\Delta_{0,\alpha}|$ and $\xi_{\alpha}$ are obtained through least squares fitting. Since $\Delta_B$ is negligible and $\Delta_A$ maps to $\Delta_C$ under mirror symmetry, we focus exclusively on $\xi_A$, which characterizes the vortex radius for sublattice $A$ along the $x$-direction, and omit the subscripts below. A key distinction between flat-band vortices and BCS vortices is presented in Fig.~\ref{fig3}(a). %\textcolor{cyan}{(Lc: Omit subscript A from here?)} 
We find that $\xi$ in the isolated flat-band vortex ($t_2=0$, blue line) remains nearly independent of $\Delta_0$. However, the finite-bandwidth case ($t_2=0.2$, $W_{fb}=0.8$, orange line) shows $\xi$ approximately inversely proportional to $\Delta_0$, a hallmark of BCS vortices. Remarkably, despite the vanishing $v_F$ in the isolated flat-band vortex, $\xi$ maintains a finite value, which is attributed to the finite quantum metric length $\bar{g}_{ab}$, as we demonstrate below. 
%This can be understood from the GL theory where the vortex radius reflects the GL coherent length near $T_c$~\cite{Tinkham2004book}. Additional details including the fitting for $\Delta_C(\mathbf{r})$ are provided in the SM.
%\textcolor{red}{LH: add figure in the SM.} 
%\textcolor{cyan}{Lc: C is connect with A via $M_{\pi/4}$, so the results along different directions will include the C results.}

%To further elucidate the role of quantum metric in governing flat-band vortex size, 
We systematically calculate the vortex size $\xi$ across temperatures $T\in[0,T_c)$. At $T=0$, $\xi$ is small and confined to a single unit cell, which is consistent with the sharply localized character of flat-band vortex states. As $T$ rises toward the critical temperature $T_c$~\footnote{Here, $T_c = 0.0515$ is calculated for the vortex-free system, with only $\pm1\%$ variation as $t_1'$ changes.}, $\xi(T)$ exhibits significant expansion [Fig.~\ref{fig3}(b)], consistent with the expected divergence at $T=T_c$. Notably, in the flat-band vortex, $\xi$ exhibits a universal dependence on larger $t_1'$ ($=0.1,0.3,0.5$) across all temperature: it decreases monotonically as $t_1'$ increases [Fig.~\ref{fig3}(b)]. This behavior is further supported by Fig.~\ref{fig3}(c), where we explicitly plot $\xi$ as a function of $t_1'$ at both zero and finite $T$. In the isolated flat-band vortex region [$t_1'\gg 0.04$, blue region in Fig.~\ref{fig3}(c)], we find that $\xi$ follows a pronounced inverse dependence on $t_1'$. Crucially, in this regime, $t_1'\gg 0.04$ only modulates the quantum metric length $\bar{g}_{ab}$ of the isolated flat-band system. This indicates a fundamental relationship between $\xi(T)$ and $\bar{g}_{ab}$.

We next compare our results with the GL theory~\cite{Law2024PRL}, where the GL coherent length $\xi_{\text{GL}}(T)$ is shown to be directly determined by the quantum metric length,
\begin{align}\label{eq:xi-det-g}
\xi_{\text{GL}}(T) = [\det{\bar{g}}]^{1/4} \tilde{T},
\end{align}
where $\tilde{T}=\sqrt{T_c/|T_c-T|}$ encodes the $T$-dependence. Our numerical results confirm the universal divergence: the vortex size $\xi(T)$ indeed scales as $\xi(T) \propto \tilde{T}$ near $T_c$. To compare with Eq.~\eqref{eq:xi-det-g} quantitatively, we extract the $T$-independent part of $\xi(T)$ by plotting $\xi(T)/\tilde{T}$ versus $[\det{\bar{g}}]^{1/4}$ in Fig.~\ref{fig3}(d). In the isolated flat-band vortex region (blue region, small $\det{\bar{g}}$), we find a linear relationship between $\xi(T)/\tilde{T}$ and $[\det{\bar{g}}]^{1/4}$ at all temperatures. %\textcolor{cyan}{(Lc: might be coincidental for low-temperature cases)} 
Thus, our results are in semi-quantitative agreement with the GL theory~\cite{Law2024PRL}. However, at small $T \ll T_c$, our results show a finite intercept from the blue and orange lines in Fig.~\ref{fig3}(d), indicating a modified scaling: $\xi(T)/\tilde{T} \approx [\det{\bar{g}}]^{1/4} + f(T)$, where the intercept $f(T)\geq 0$ vanishes only as $T \to T_c$. In general, this implies a lower bound for $\xi(T)/\tilde{T}$ set by $[\det{\bar{g}}]^{1/4}$. This conclusion remains valid for $\xi$ extracted along other directions at high $T$ [see SM~\cite{SP}]. However, beyond the isolated flat-band vortex limit (large $\det \bar{g}$), where the dispersive bands become increasingly relevant, the linear scaling relation breaks down. This underscores the distinct physics of the non-isolated flat-band vortices.

\textit{\color{blue}Conclusions and discussions.--}
In conclusion, we have established vortices in flat-band superconductors as direct probes of quantum geometry effect. Our results reveal fundamental distinctions from BCS theory. By mapping the crossover from isolated flat-band vortices to BCS vortices, we find unique properties in flat-band vortex, including: (i) only order of one low-energy bound state appears; (ii) sharply localized flat-band vortex states exhibit atomic-scale spatial confinement, with their spread and energy governed by the quantum metric length $\text{Tr}[\bar{g}]$; 
%\textcolor{cyan}{(Lc: is quantum metric length defined $(\text{Tr}[\bar{g}])^{1/2}$ or $[\det{\bar{g}}]^{1/4}$?)} 
and (iii) the flat-band vortex size diverges as $\xi(T) \approx \sqrt{T_c/|T_c-T|}\xi_0$ 
%\textcolor{cyan}{(Lc: $\propto \to \approx$)} 
near $T_c$, where $\xi_0$ linearly depends on $[\det{\bar{g}}]^{1/4}$. In general, the flat-band vortex size $\xi(T)$ deviations from the GL theory as $[\det{\bar{g}}]^{1/4}$ provides a lower bound. This is supported by experimental measurements in twisted bilayer graphene showing $\xi_{\text{expt}}\approx 55$ nm~\cite{Bockrath2023Nat}, while the GL theory considering only quantum metric contributions yields $\xi_{\text{GL}}\approx 30$ nm~\cite{Law2024PRL}. 
Our theory bridges quantum geometry and vortex physics, providing a framework to experimentally explore flat-band superconductivity (e.g,~via scanning tunneling microscopy).

Our work opens several promising avenues for future research. First, generalization to other isolated flat-band superconducting systems, like kagome and checkerboard lattices could reveal universal versus material-specific quantum metric effects. Second, the Lieb lattice model employed in this work exhibits intrinsic structural anisotropy, but the flat-band superconductivity display near-isotropic vortex properties. Introducing controlled anisotropy could probe new quantum metric effect via directional responses in vortex confinement and dynamics. Third, the observed lower bounds for flat-band vortex size and spread of bound states require a rigorous analytical framework for deeper understanding. Moreover, flat-band superconducting materials like Mo$_5$Si$_{3-x}$P$_x$~\cite{Khasanov2024NatCom} and LaRu$_3$Si$_2$~\cite{DJunZe2025arXiv} warrant combined theoretical and experimental investigations of flat-band superconductivity.

\textit{Acknowledgments.--}
We thank G.~D.~Jiang, P\"aivi T\"orm\"a, S.~B.~Zhang, Z.~Wang and H.~K.~Jin for helpful discussions. 
L.H.H. is supported by the start-up of Zhejiang University and the Fundamental Research Funds for the Central Universities (Grant No. 226-2024-00068).

\bibliographystyle{apsrev4-1}
\bibliography{ref}
\end{document}